\documentclass[preprint,times]{aastex631}
\usepackage{cancel}
\usepackage{amsmath}
\usepackage{amsfonts}
\usepackage{amssymb}
\usepackage{empheq}
\usepackage{natbib}
\usepackage{tikz}
\usepackage{ulem}

\revised{2021 Oct 18}

\shorttitle{Cometary Activity Beyond the Planets}
\shortauthors{N. Bouziani \& D. Jewitt}

\begin{document}

\title{COMETARY ACTIVITY BEYOND THE PLANETS}

\author[0000-0002-3805-7120]{Naceur Bouziani}
\affiliation{Center of Research in Astronomy, Astrophysics and Geophysics \\
	Route de l'Observatoire, 16340 Bouzareah, Algiers, Algeria}

\author{David Jewitt}
\affiliation{Department of Earth, Planetary and Space Sciences, UCLA \\ 595 Charles Young Drive East, Los Angeles, CA 90095-1567, USA}
\affiliation{Department of Physics and Astronomy, UCLA \\ 430 Portola Plaza, Box 951547, Los Angeles, CA 90095-1547, USA}


\begin{abstract}
Recent observations show activity in long-period comet C/2017 K2 at  heliocentric distances beyond the orbit of Uranus. With this as motivation, we constructed a simple model that takes a detailed account of gas transport modes and simulates the  time-dependent sublimation of super-volatile ice from beneath a porous mantle on an incoming cometary nucleus. 
The model reveals a localized increase in carbon monoxide (CO) sublimation close to heliocentric distance $r_H$ = $150$ AU (local blackbody temperature  $\sim$23 K), followed by a plateau and then a slow increase in  activity towards smaller distances.  This localized increase occurs as heat transport in the nucleus transitions between two regimes characterized by the rising temperature of the CO front at larger distances and nearly isothermal CO at smaller distances.  As this transition is a general property of sublimation through a porous mantle, we predict that future observations of sufficient sensitivity will show that in-bound comets (and interstellar interlopers) will exhibit activity at distances far beyond the planetary region of the solar system.
\end{abstract}


\email{naceur.bouziani@craag.edu.dz}

\section{Introduction} \label{sec:intro}

It has long been known that some comets are active far beyond the range of heliocentric distances over which water ice can sublimate, the latter corresponding roughly to the region interior to Jupiter's orbit (semimajor axis 5 AU).  The interpretation of activity in distant comets is complicated.  Ices more volatile than water (e.g.~CO, CO$_2$) may be involved, as might the exothermic crystallization of amorphous water ice, at least at distances $r_H \lesssim$ 12 AU \citep{2009AJ....137.4296J, 2015SSRv..197..271G}.  In comets observed  after perihelion, the origin of distant activity is further complicated by the slow conduction into the interior of heat acquired at perihelion, resulting in the delayed activation of sub-surface volatiles and driving distant outbursts (e.g.~\cite{1992A&A...258L...9P}).  \\

Significantly, in  long-period comets that are active at large distances while still inbound to perihelion, the conduction of perihelion heat can play no role.  The long-period comet C/2017 K2 (PANSTARRS), which was discovered active pre-perihelion at  heliocentric distance  $r_H = 16$ AU, falls into this category.  Observations reveal a spherical coma of large ($\sim$0.1 - 1 mm) size grains, slowly ejected (speeds $\sim$ 4  m s$^{-1}$) close to steady-state from a nucleus no more than 9 km in radius, \cite{2017ApJ...847L..19J, 2019AJ....157...65J}.   Coma was present in prediscovery observations at 23.8 AU,  a linear extrapolation of photometry indicates that activity began at $r_H \sim$ 26 AU, \citep{2019AJ....157...65J} while more detailed considerations indicate that C/2017 K2 was active even at $\sim$35 AU \citep{2021AJ....161..188J}.  Comet C/2010 U3 (Boattini)  likewise  possessed a coma at 25.8 AU pre-perihelion \citep{2019AJ....157..162H}, while C/2014 B1 (Schwartz) was also active far beyond the water sublimation zone with a coma of large, slowly-moving grains, but observed only starting at 11.9 AU, pre-perihelion \citep{2019AJ....157..103J}.  Recently, comet C/2014 UN271 (Bernardinelli-Bernstein) was reported to be active at 23.8  AU \citep{2021ATel14759....1F} and may have been active at much larger distances.   \\

Free sublimation of exposed super-volatile ices  provides an initially attractive but ultimately unsatisfactory explanation for the distant activity. Gas drag forces  produced by free-sublimation of CO  in equilibrium with sunlight at distances $>$ 15 AU  are orders of magnitude too small to eject 100 $\mu$m sized particles against the gravity of the nucleus \citep{2019AJ....157...65J}.  Neither could smaller particles  be ejected owing to  inter-particle (van der Waals) cohesive forces naturally between grains.  Taken at face value, this ``cohesion bottleneck'' prevents the ejection of particles of any size in distant comets, as  noted by \citet{2015A&A...583A..12G}.  However, sublimation in a confined space, for example beneath a porous refractory mantle, offers the possibility of building the pressure to values much higher than reached in free sublimation to space.  \cite{2020A&A...636L...3F} have suggested that sublimation occurs within centimeter-sized ``pebbles'' whose small but finite strength allows gas pressure to build up to values sufficient to eject dust by gas drag.  Activity driven by the crystallization of amorphous water ice has also been proposed as a mechanism but, at $\ge$ 15 AU, temperatures are too low for this process to occur.   \\

We here explore solutions to the bottleneck problem involving the build-up of pressure by the sublimation of sub-surface ice, taking careful account of the processes by which heat and gas transport occur.  In so doing, we find that cometary activity is possible at extraordinarily large distances, even far beyond the planetary region of the solar system.  We present the model in Section (\ref{sec:nucleus}) and the results in Section (\ref{results}).

\section{The Model } \label{sec:nucleus}

\begin{figure}[ht!]
	  \plotone{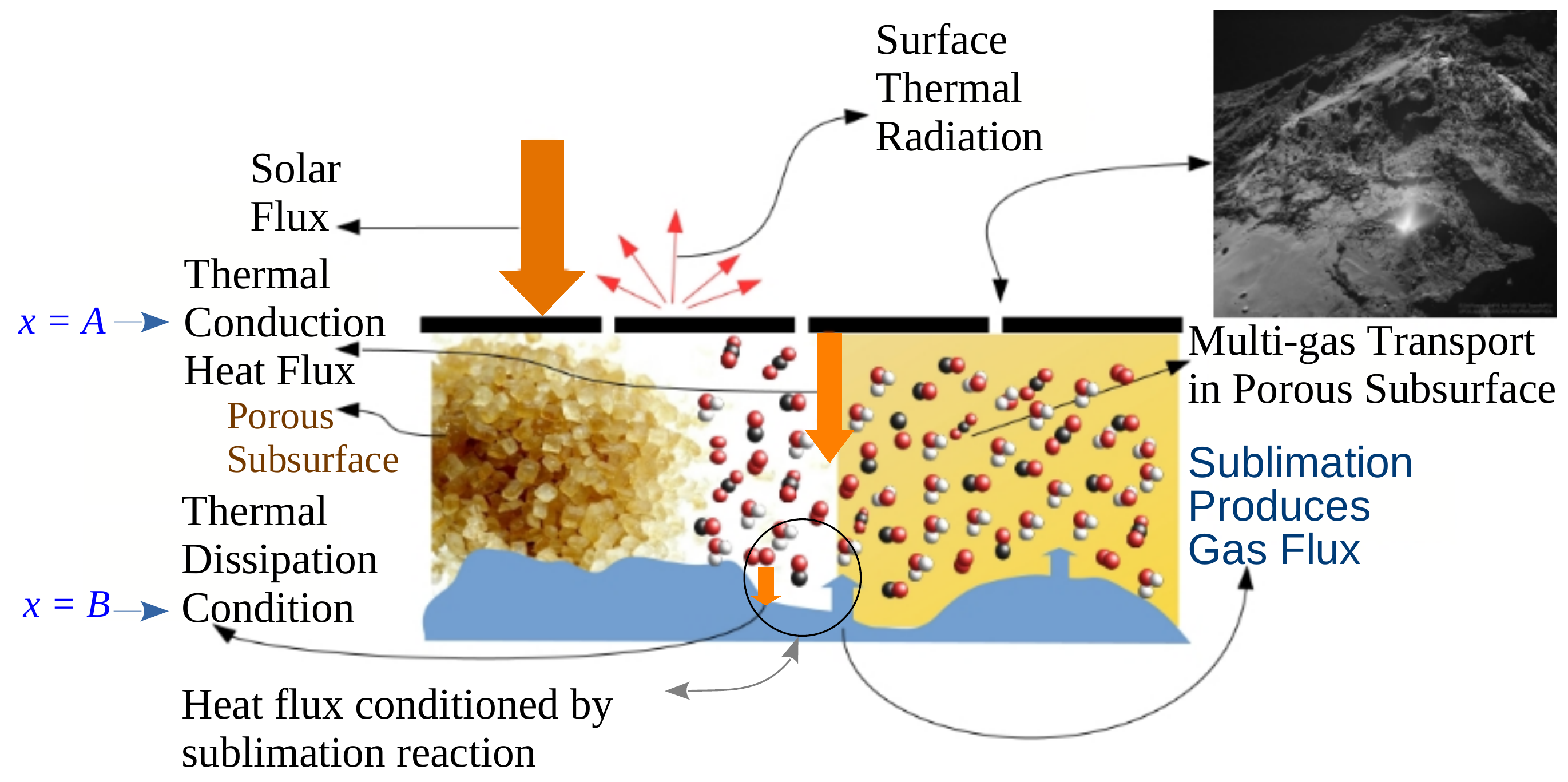}
	 \caption{Schematic of the model physics\label{Fig1}.  The photo on the right is from the ESA Rosetta Mission to 67P/Churyumov-Gerasimenko (Osiris Team), the one on the left is sugar powder from J.~C.~M. \url{https://stock.adobe.com/contributor/265923/jcm}}
\end{figure}

The physical model we present is a refinement of Whipple's (1950) icy conglomerate model, enhanced to include the presence of a non-volatile, low conductivity porous mantle (see Figure \ref{Fig1}). In active comets, such a ``rubble mantle'' is produced by fallback of suborbital particles and by volatile loss from the strongly heated surface.  In inactive comets, for example those in the Oort cloud and Kuiper belt storage reservoirs, an irradiation mantle can grow as  pure ices are rendered non-volatile by  prolonged exposure to cosmic rays (\cite{1997ApJ...475..144K}, \cite{2019SciAdv..5..eaaw5841}, \cite{2019EPSC...13..914G}). In our model, we assume that a non-volatile  mantle of variable thickness and porosity overlies  CO ice. \\

In equilibrium with sunlight, the energy balance averaged over the surface of a spherical, isothermal nucleus is written as 

\begin{equation}\label{key1}
\frac{F_\textsubscript{\(\odot\)} \left(1-A_{B}\right)}{r_{H}^{2}}+ 4 \sigma \epsilon_{IR} T_{OC}^{4}= 
4 \sigma  \epsilon_{IR} T^{4} + 4  k(T) \frac{\partial{T}}{\partial{x}}
\end{equation}

The first term on the left of Equation (\ref{key1}), in which $F_{\odot}$ is the solar constant, $r_H$ is the heliocentric distance in AU and $A_B$ is the bond albedo, represents the absorbed solar power.  The second term on the left, only important at the largest heliocentric distances, accounts for heating of the nucleus from other external sources. These include the cosmic microwave radiation background and the integrated light of the stars, as represented by an effective temperature $T_{OC}$ = 10 K.  The emissivity of the nucleus is $\epsilon_{IR}$ while $\sigma$ is the Stefan-Boltzmann constant.  On the right, the first term is the flux of radiation from the nucleus surface at temperature, $T$, while the second term is the conducted flux.  There is no contribution from sublimation since the surface layers are assumed to be volatile-free.  \\

The conductivity in Equation (\ref{key1}), $k(T)$, is written as in \cite{1977Moon...17..359M}

\begin{equation}\label{key2}
k(T)=A+B(4\sigma \epsilon_{IR}T^{3})
\end{equation} 

\noindent where the first term, $A$, is the effective thermal conductivity of the porous mantle and second term accounts for radiative conductivity.  $B$ is an empirical attenuation factor for radiative conduction of void and adjacent grain surfaces within the porous medium.  \\

As shown schematically in Figure (\ref{Fig1}),  the ice is not on the surface as in Whipple's model but hidden under a mantle. Energy propagated through the mantle to deeper layers can trigger the sublimation of buried ices.  At the ice-mantle interface, the conducted energy flux sublimates ice at rate 
 
\begin{equation}\label{key3}
      k(T) \frac{\partial{T}}{\partial{x}}= E_{CO} f_{CO}
\end{equation}

\noindent in which we ignore heat conducted into the ``cold core'' beneath the sublimation front.  A justification of this neglect is given in the Appendix.  \\

Here, the sublimation is entirely controlled by the conducted heat; a non-equilibrium process.  The generated gas is also subject to another non-equilibrium process caused by the pressure gradient.  The heat is driven down the temperature gradient towards the ice front, $B$ in Figure (\ref{Fig1}), while the gas is driven up towards the surface of the comet, $A$, by the pressure gradient established in the porous medium. In this simple comet model the reverse flow of molecules is not favored, so condensation is somewhat unlikely and is neglected for simplicity. The sublimation flux that has been calculated in \cite{1961JGR....66.3033W} and in \cite{1971P&SS...19.1229D} approximates the sublimation flux in our case too. Otherwise, we should model the condensation process and calculate the net flux

\begin{equation}\label{key9}
f_{CO}  =  P(T) \sqrt{\frac{1}{2\pi  k_{B}  T m}}
\end{equation}

Practically, $f_{CO}$ is an upper limit to the sublimated flux because of the possibility of back-flow and sticking of molecules reflected from particles above the ice front.   This is the necessary Neumann boundary condition.  At all levels above the ice front, we calculate the net flux, $J_{CO}$, formed as molecules diffuse through the porous medium.  \\

In equation (\ref{key9}), $P(T)$ is the equilibrium pressure of the $CO$ gas over the $CO$ ice as function of temperature. $P(T)$ contains valuable information about both the gas and  the ice structure.  We consider sublimation and condensation as facets of a single process which is a first order phase transition phenomenon. It is worthwhile to invoke the chemical reaction formalism to address it. Sublimation is  considered as the direct process while condensation is the reverse process; analogous to the direct and reverse arrows between reactants and products in chemical reaction theory. Usually, experiments are conducted in closed thermodynamic equilibrium to characterize the chemical reaction and find the equilibrium constant called $K(T)$. We  found that for pure ice, the temperature dependence of $P(T)$ varies just like $K(T)$ in the language of chemical reactions. The resulting pressure relation resembles  the Arrhenius equation  (\ref{key11}). Physically, the similarity arises because in a chemical reaction we consider the breakage of bonds within molecules while, in sublimation, the breaking bonds are those between molecules.  For this reason, the  activation energies $E_i$ in the sublimation case are  small compared to the energies involved in chemical reactions.  \\

\begin{figure}[ht!]
	\plotone{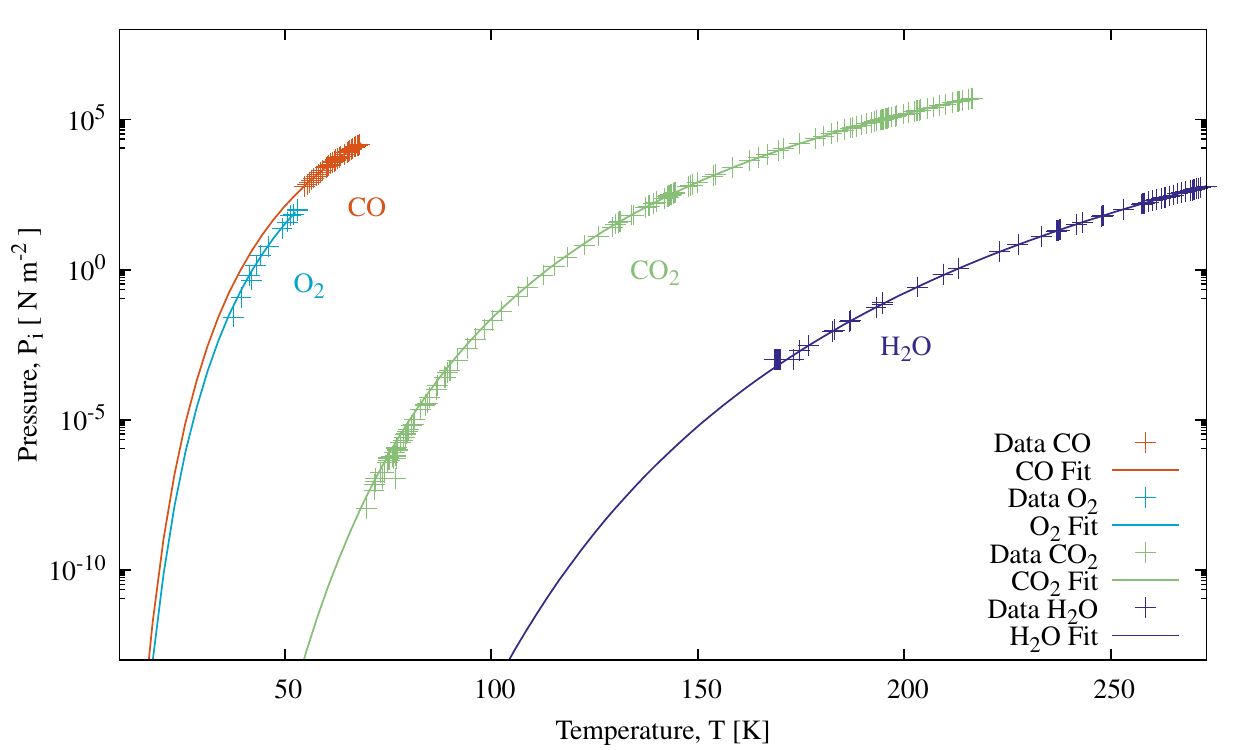}
	\caption{Sublimation pressures from Fray \& Schmitt~(2009) fitted by the relation $P_{i}(T) = A_{i} e^{- E_{i} / RT}$ to determine the activation energy, $E_{i}$ (J mol$^{-1}$), and the frequency factor, $A_{i}$. The resulting numerical values are listed in Table (\ref{tab1}). 	\label{Fig2}}
\end{figure}

We fitted the sublimation data from \cite{2009P&SS...57.2053F} using

\begin{equation}
\label{key11}
P(T)  =  A_{CO} e^{-{E_{CO}}/{RT}}
\end{equation}

\noindent where $R$ is the molar gas constant, $E_{CO}$ is the activation energy of $CO$  sublimation, and $A_{CO}$ is a constant, using a non-linear Levenberg Marquardt technique.  The data and the fits are shown in Figure  (\ref{Fig2}). 
Equation (\ref{key11}) provides a better fit to the data than the polynomials used by \cite{2009P&SS...57.2053F}. We succeeded in inferring the activation energy of $CO$ with an exceptional accuracy, see the energies actually measured using  various dedicated experimental methods and we did this independently and in a temperature range not included in the data  \citep{2014A&A...566A..27L}.  \\

Our model is governed by two coupled partial differential equations, one for heat transport 

\begin{equation}
\rho  c_{p}(T) \frac{\partial{T}}{\partial{t}} = \frac{\partial }{\partial x}\left(k(T)\frac{\partial{T}}{\partial{x}}\right)\\\label{key5}
\end{equation}

and one for mass transport
 
 \begin{equation}
\label{key6}
\frac{\partial n}{\partial t}~~~~~=~~~\epsilon\frac{\partial }{\partial x}\left(D_{K}\frac{\partial n}{\partial x}+\frac{\kappa n}{\mu} \frac{\partial p}{\partial x}\right).
\end{equation}

In Equation (\ref{key5}) $\rho$ is the density and $k(T)$ is the conductivity at temperature, $T$, while $t$ is the time and $x$ is the depth in the nucleus, and $c_{p}(T)$ (see equations  \ref{newkeycp2} and  \ref{newkeycp1}) is the heat capacity of the mantle. In Equation (\ref{key6}), $D_{K}=\frac{r_{p}}{3}V_{th}$ is the Knudsen diffusion coefficient, $V_{th}=\sqrt{\frac{8k_{B}T}{\pi m}}$ is the thermal speed, $\kappa=\frac{r_{p}^{2}}{32}$  the permeability of the porous medium  \citep{cussler2009diffusion}, and $\mu = \frac{\alpha}{\pi^{3/2}}\frac{\sqrt{k_{B}mT}}{\sigma^{2}}$ is the dynamic viscosity.  Other quantities are  $k_{B}$, the Boltzmann constant, $\alpha$, a numerical constant on the order of 1, $T$, the temperature, $\sigma$  the CO molecule's collision diameter and $m$ the CO molecular mass, $p$, the CO gas pressure $p(n,T)= nk_{B}T$, $n$ is the molecular density of CO; $r_{p}$ the pore radius, $\epsilon$ is the porosity.

\begin{figure}[ht!]
	\plotone{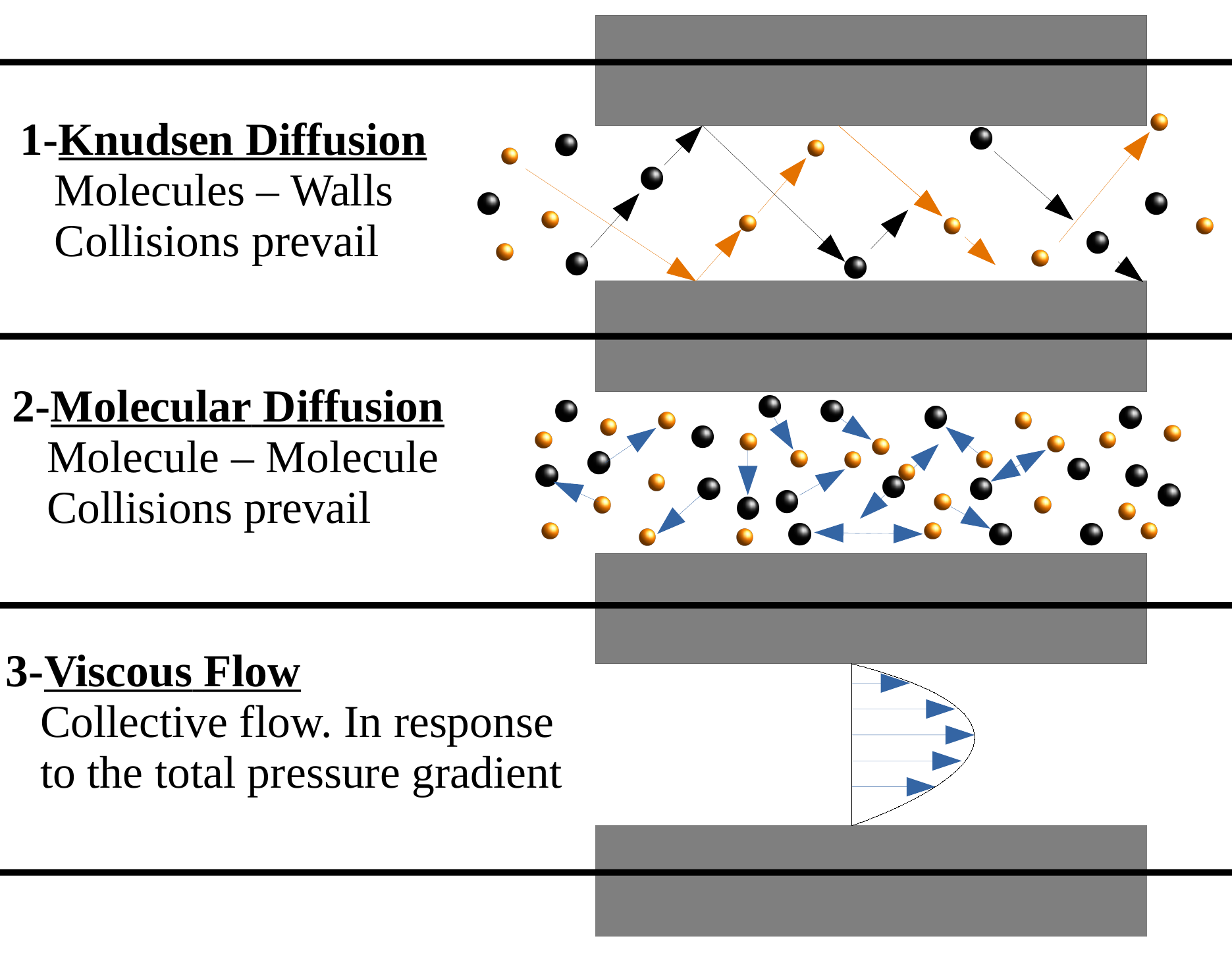}
	\caption{Schematic showing different molecular transport modes. In the porous mantle, gas transport is hindered by the walls encountered by the molecules during their journey. Panels (1), (2) and (3) show the three gas transport mechanisms within the porous mantle. This figure is adapted from \url{https://tinyurl.com/y3xyj9ss}.\label{Fig3}}
\end{figure}

The equations are coupled by the transport coefficients. Water ice acts as  a mineral at the low temperatures found in the outer solar system, so we neglect the H$_2$O gas flux and focus only on the CO flux using a simplified version of the model described in \cite{1998ApJ...499..463B}. Gas produced by sublimation passes through the porous mantle towards the comet surface. Figure (\ref{Fig3}) illustrates the various modeled mechanisms of gas transport inside the porous mantle. This model takes into account both microscopic and macroscopic behavior and also treats the interaction between the gas and the walls. It provides a continuous physical modeling of the coupling between mechanisms - from discontinuous collisions to the collective macroscopic behavior of continuous matter - and thus for the prediction of the mass flow rate and pressure distribution with reasonable accuracy over the widest range of gas pressures. The model depends only on molecular data and fundamental parameters such as temperature, composition and total pressure. This approach allows us to address the thermal and dynamical coupling of comets in a continuous and physical way.  Table (\ref{tab1}) summarizes parameters used in the numerical simulations. \\

We assume that the dust mantle is composed of a mixture of refractory minerals and water ice. To estimate the heat capacity of the ice, we use the heat capacity $c_{p}$ (J kg$^{-1}$ K$^{-1}$)deduced from the data provided in \cite{2004A&A...416..187S}.

\begin{equation}\label{newkeycp2}
\begin{split} 
c_{p}(ice)= 7.73 T \left(1-e^{-1.263 \times 10^{-3} T^{2}}\right)
\left(1 + e^{-3\sqrt{T}} \times 8.47 T^{6} + 2.0825 \times 10^{-7}T^{4}e^{-4.97 \times 10^{2}T} \right)
\end{split} 
\end{equation}
 
For the refractory part of the mantle we use the data for Enstatite found in \cite{krupka1985low} and we find that the Shulman equation gives an excellent representation of the data at both high and low temperatures;

\begin{equation}\label{newkeycp1}
\begin{split} 
c_{p}(dust)= 2.03 T \left(1-e^{-2.5 \times 10^{-4} T^{2}}\right)
\left(1 + e^{-8.53\times 10^{-1}\sqrt{T}} \times 1.24\times 10^{-09} T^{6} -3.69 \times 10^{6}T^{4}e^{-4.76 \times 10^{5}T} \right).
\end{split} 
\end{equation}

\begin{figure}[ht!]
	\plotone{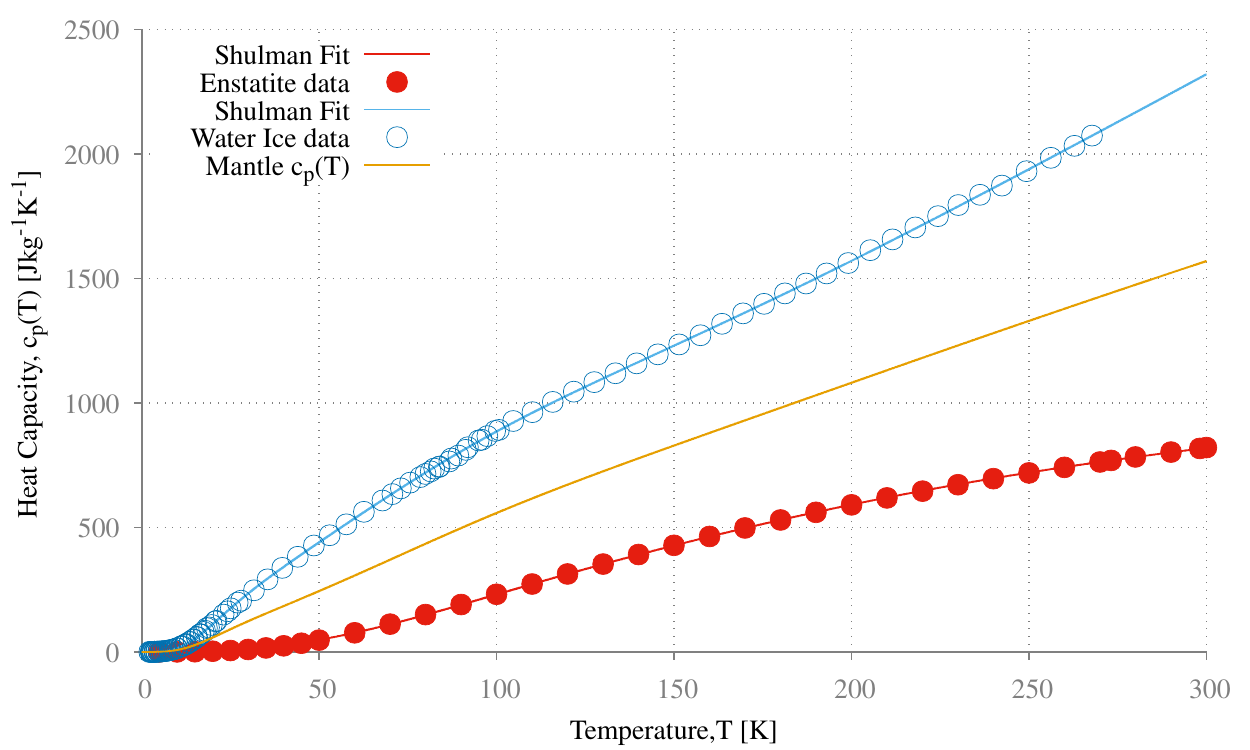}
	\caption{Heat capacity of the mantle as a function of temperature, $c_{p}(T)$ in (J kg$^{-1}$ K$^{-1}$)
	\label{Fig4}}
\end{figure}

\noindent  We assume a water ice/dust ratio = 1, and thus take the specific heat capacity of the mixture as the average of equations (\ref{newkeycp1}) and (\ref{newkeycp2}). Figure (\ref{Fig4}) shows how the heat capacity of the mantle  depends on the temperature.
\\
 
We solved Equations (\ref{key5} \& \ref{key6}) as a function of time, following the motion of comet C/2017 K2 upon its approach  to the Sun from large heliocentric distances.  For this purpose, we used the barycentric orbital elements from \cite{2018A&A...615A.170K} while noting that the current Sun-centered orbit  from NASA's Horizons ephemeris software\footnote{\url{http://ssd.jpl.nasa.gov/horizons.cgi}} is slightly hyperbolic (eccentricity $e$ = 1.0004, vs.~$e$ = 0.9999 from \cite{2018A&A...615A.170K}).  This difference between the barycentric and Sun-centered elements is negligible for our purposes.  
The model solves the heat and gas transport  equations  in a coupled manner. As it approaches the Sun, the surface temperature of C/2017 K2 increases from the limiting interstellar value of 10 K to higher temperatures, creating a temperature gradient within the mantle, and driving  conduction towards the buried ice layer. Heating of the ice triggers sublimation, producing a pressure gradient and resulting in a diffusive flow of gas towards the comet's surface, eventually leading to escape and the expulsion of embedded dust particles through gas drag.  The downward conduction of heat and the upward transport of sublimated gas molecules are characterized by distinct and different timescales, each dependent on the thermal and micro-physical properties of the material, as we will discuss in Section (\ref{retention-time}). The boundary condition (Equation \ref{key1}) is time-dependent because the orbital position is a function of time.

\section{Results \& Discussion }
\label{results}

\subsection{The Activity Mechanism}
\label{activity-mechanism}
\begin{figure}[ht!]
	\plotone{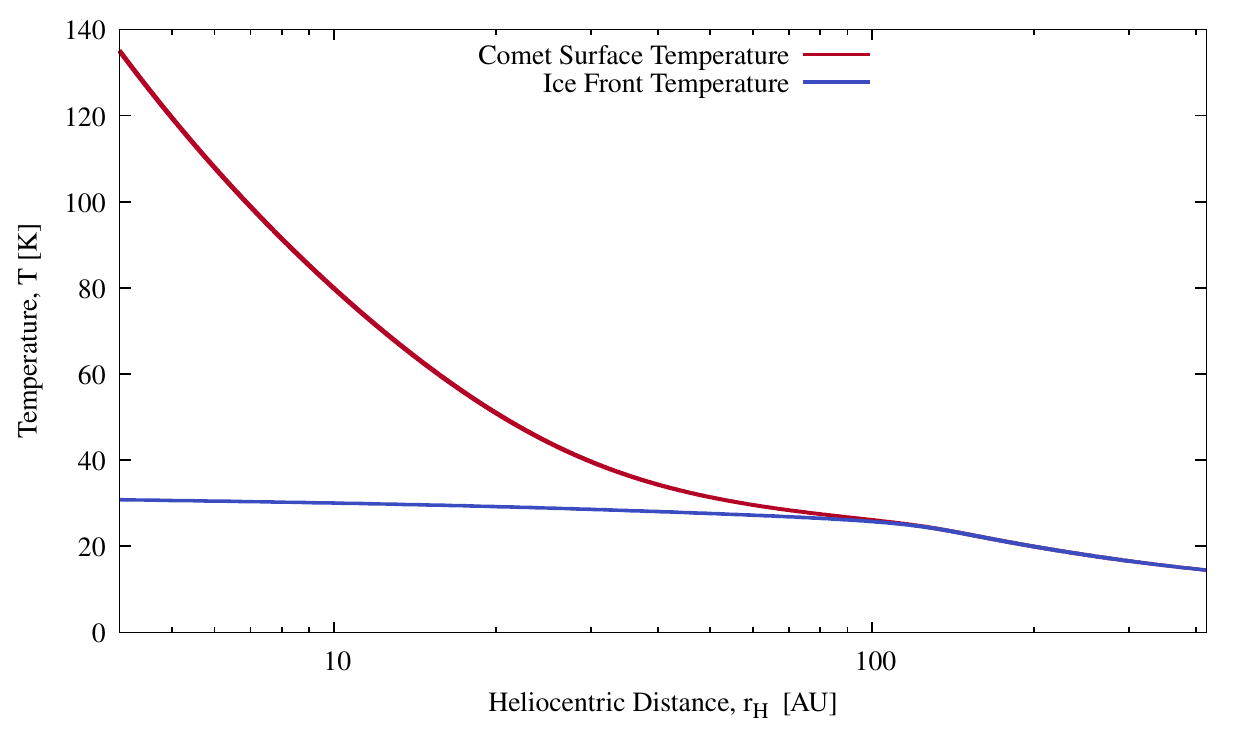}
	\caption{Nucleus surface (red line) \& buried ice front (blue line) temperatures as a function of heliocentric distance, for  in-falling comet C/2017 K2.  In this illustrative example the effective thermal conductivity is $k(T)= 10^{-2}$ W m$^{-1}$ K$^{-1}$, pore radius is $r_p = 10^{-4}$ m, 30\% of the flux is allowed to escape the mantle, and the mantle thickness is 1 m. \label{Fig5}}
\end{figure}

Figure (\ref{Fig5})  shows that the surface temperature increases, as expected from radiative equilibrium, in proportion to $r_H^{-1/2}$.  On the other hand, the temperature at the ice-mantle boundary at first rises as $r_H$ decreases (we refer to this region as Regime 1) but becomes nearly constant at $r_H \lesssim$ 150 AU (Regime 2).  This flattening occurs because the heat conducted to the ice-mantle boundary in Region 2 is almost entirely used to break intermolecular bonds in the ice, leaving little to raise the temperature.  These qualitatively different temperature vs.~distance trends were first noted for H$_{2}$O by \cite{1977Moon...17..359M} for Comet Kohoutek (1973f) and later by \cite{1988Icar...74..272P}. The model confirms this behavior and extends it to much larger heliocentric distances in more volatile ices.  For convenience, in the remainder of this paper we refer to the junction between Region 1 and Region 2 as the ``Mendis point''. The CO flux coming out from the comet mantle is given by Equation (\ref{key6}). In order to better understand the orbital behavior of this flux as the comet approaches the sun, we solve the coupled equations (\ref{key5} \& \ref{key6}) and focus on the averaged flux $J_{CO}$ by estimating the difference in the value of a variables between the surface of the comet $A$ and the base of the mantle $B$, this approximation is emphasized next in section (\ref{retention-time}); doing so, we deliberately ignore the local variation of this flux within the mantle.  The flux $J_{CO}$ is

\begin{equation}
J_{CO}~~=~~\epsilon \left(D_{K}\frac{d n}{d x}+\frac{\kappa n}{\mu} \frac{d p(n,T)}{d x}\right)\label{key6bis1}.
\end{equation}


Equation (\ref{key6bis1}) shows how Knudsen free diffusion combines with collective viscous flow (also called Hagen-Poiseuille flow in fluid dynamics). Although these two flows are additive, they are not independent of each other. They are coupled through the pre-gradient coefficients, thanks to the microscopic approach of the model which integrates the walls in the collision from the very beginning. \\
  
\begin{figure}[ht!]
	\plotone{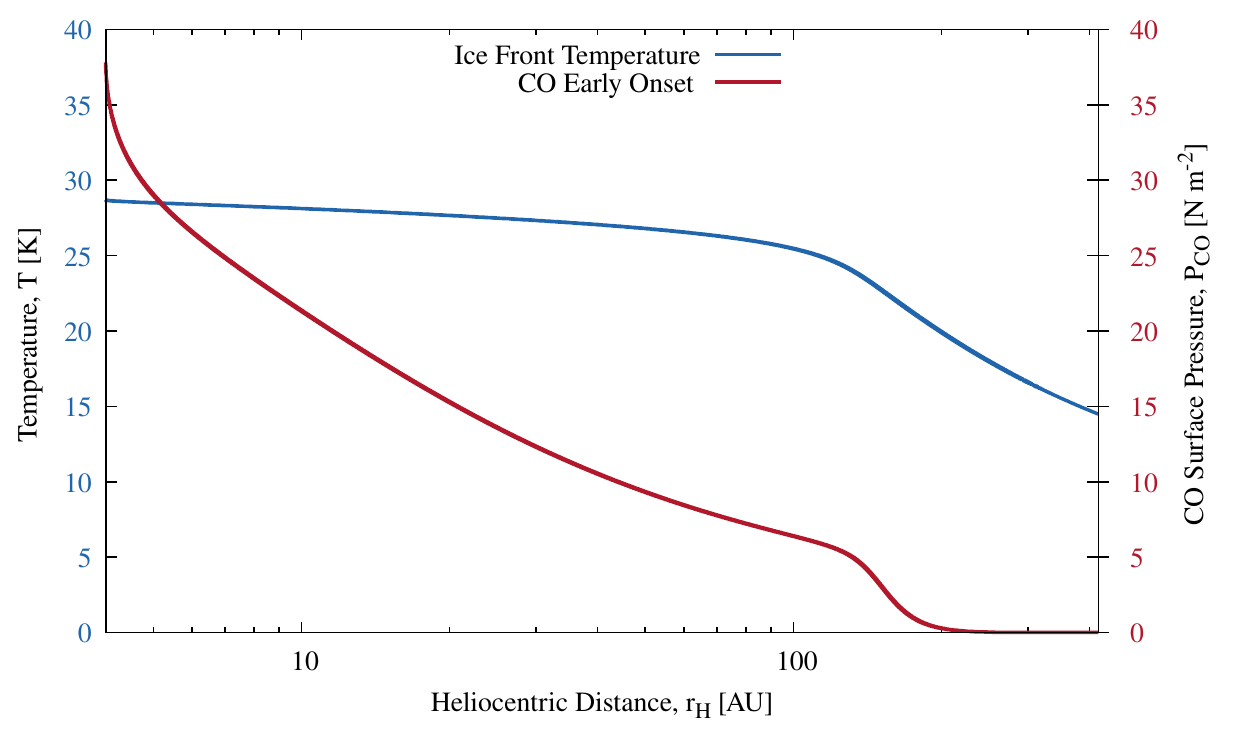}
	\caption{ CO ice front temperature (blue line) and pressure just bellow the surface, $p_{CO}$ (red line), the latter showing the Mendis point at about 150AU.  This simulation  assumes thermal conductivity  $k(T)= 10^{-4}$ W m$^{-1}$ K$^{-1}$ and a mantle thickness of 1 m. The comet surface pores are 10\% smaller than the mantle pores which have $r_p$ = 1.15$\times10^{-4}$ m. 
\label{Fig6}}
\end{figure}

\begin{figure}[ht!]
	\plotone{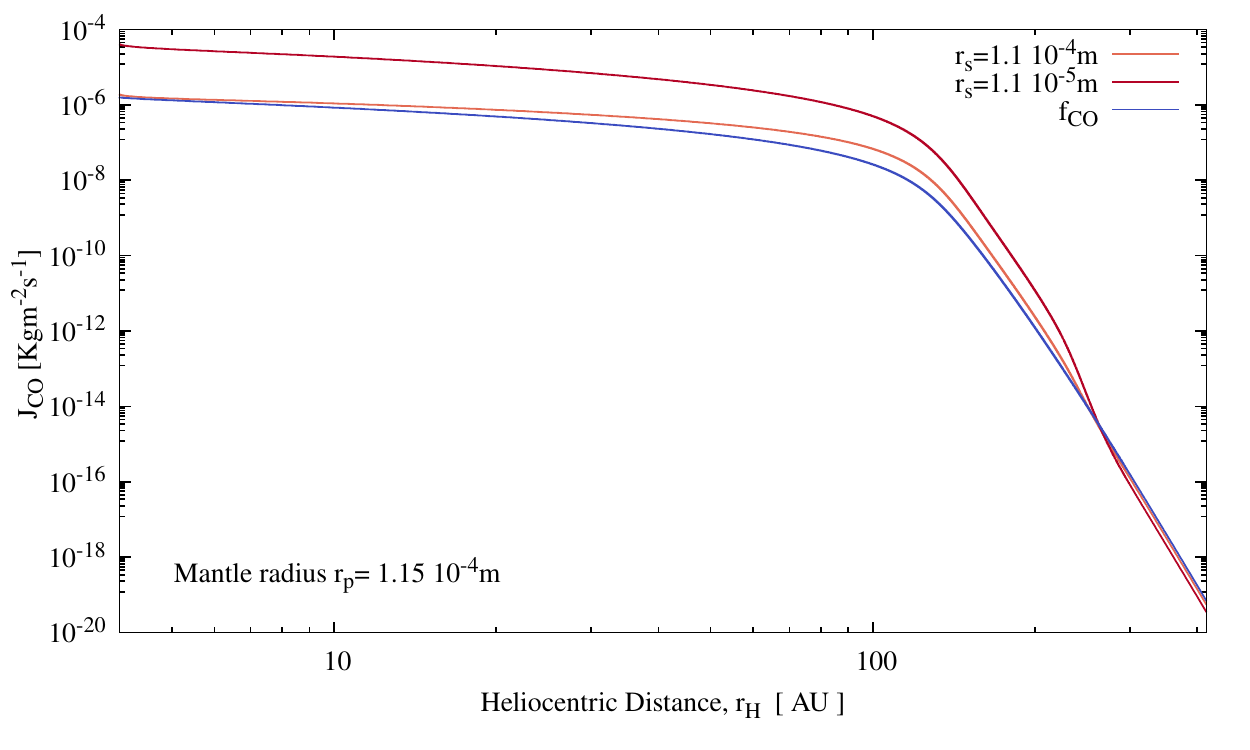}
	\caption{ CO flux $J_{CO}$ as a function of heliocentric distance for two surface pore radii, $r_s$. Mantle pore radius is $r_p $= 1.15$\times 10^{-4}$ m.  Assumed thermal conductivity is  $k(T)$= 10$^{-3}$  W m$^{-1}$ K$^{-1}$; the mantle thickness is 1 m.  \label{Fig7}}
\end{figure}

\begin{figure}[ht!]
	\plotone{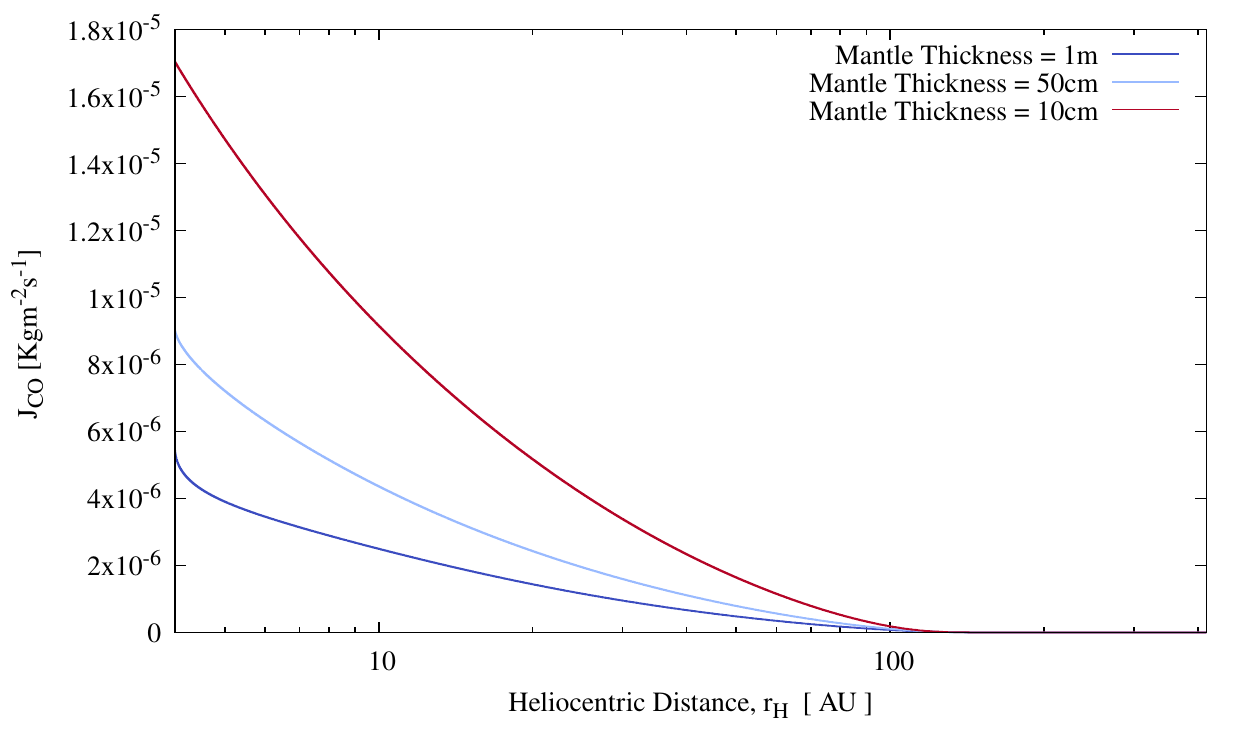}
	\caption{Dependence of the CO flux $J_{CO}$ on  heliocentric distance and  the assumed mantle thickness. 
	\label{Fig8}}
\end{figure}

Figure (\ref{Fig6}) shows the CO gas pressure just below the surface, $p_{CO}$, as a function of $r_H$. We find a prominent  local bump  in $p_{CO}$  at $r_H \approx~ 150$ AU, corresponding to the Mendis point.  Further exploration shows that the location and amplitude of the $J_{CO}$ bump  depend on several physical parameters, including the assumed active surface fraction of the mantle (the comet surface structure) and the mantle thickness.
For example,  Figure (\ref{Fig7}) shows that the amplitude of the Mendis point bump increases as the mantle active surface fraction decreases, because a lower surface pore radius, $r_{s}$, increases the gas speed and allows the build up of higher pressures that in turn speed up even more the flux (viscous mode is more rapid). Figure (\ref{Fig8}) shows as expected that flux $J_{CO}$  increases as the mantle thickness decreases; while thinner mantles provide less impedance to the flow of gas, they  also allow higher temperatures, and so higher pressures, at the ice sublimation front. \\

\begin{figure}[ht!]
	\plotone{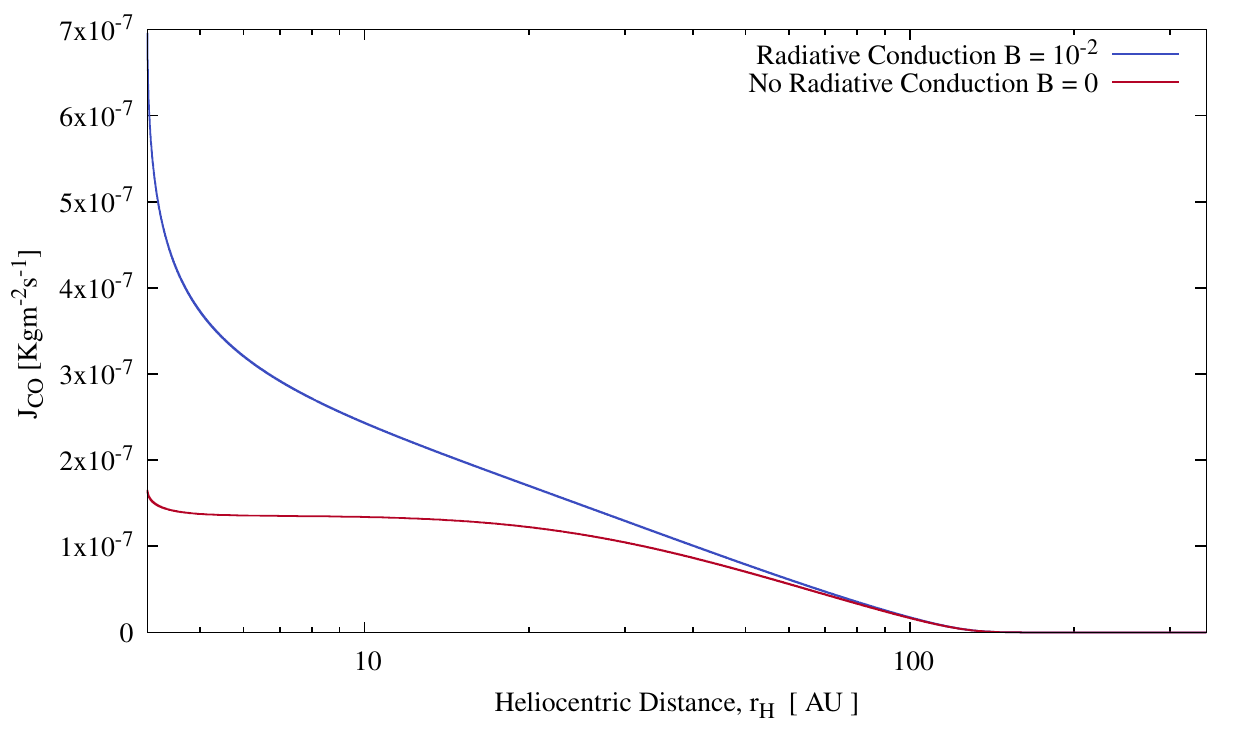}
	\caption{ Effective of radiative conduction.  Red curve shows the $J_{CO}$ flux when thermal conductivity includes a radiative contribution as described in the text. 
	 Blue curve shows the surface pressure calculated without a radiative contribution. Here mantle thickness is 5m. All  other parameters are held fixed between the models.
	 \label{Fig9}}
\end{figure}

The steady increase in $J_{CO}$ after the 150 AU Mendis point becomes an abrupt rise as the comet approaches the Sun (c.f.~Figure \ref{Fig9}). Part of this strong increase at small $r_H$ is due to the rising contribution of the radiative term to the effective thermal conductivity of the mantle. To  demonstrate this behavior, we show in Figure (\ref{Fig9}) the effect of arbitrarily removing the radiative term by setting $B$ = 0 in  Equation  (\ref{key2}).    In this model with a 5m thick mantle, the CO flux $J_{CO}$ at 4 AU falls from $6.4\times 10^{-7}$ kg m$^{-2}$ s$^{-1}$ (blue curve, including radiative term) to $\approx$ $1.2\times 10^{-7}$ kg m$^{-2}$ s$^{-1}$ (red curve, no radiative term). 
 Figures (\ref{Fig6}, \ref{Fig7}, \ref{Fig8} \& \ref{Fig9}) reveal that, depending on assumed parameters of the mantle,  radiative conduction  begins to be significant  around $50$ AU, reaches its maximum near  perihelion but has no effect at distances corresponding to the CO Mendis point.\\

\begin{figure}[ht!]
	\plotone{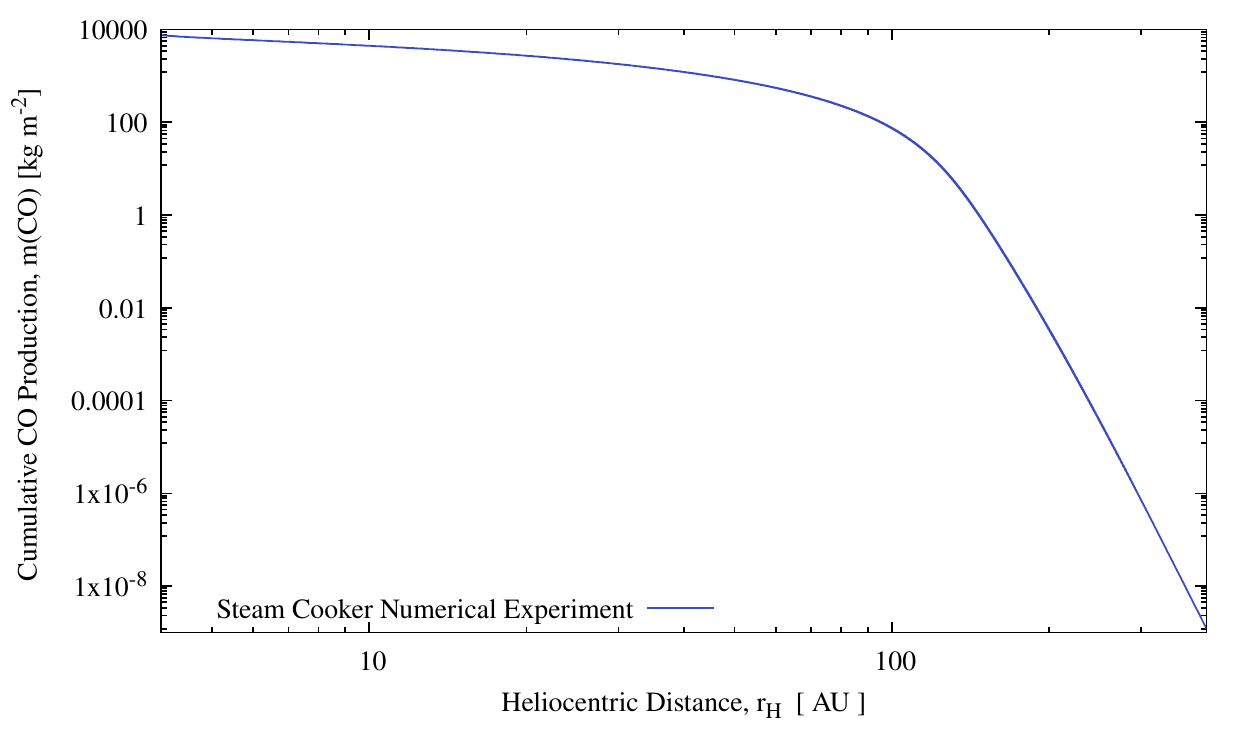}
	\caption{Extreme case of a totally sealed surface (``steam cooker'' approximation). This shows the cumulative CO gas mass production in (kg m$^{-2}$). 	For this numerical experiment we used 
	$k(T)= 10^{-2}$ W m$^{-1}$ K$^{-1}$ and the mantle thickness is 1 m. The curve is independent of the mantle structure. 
	\label{Fig10}}
\end{figure}

Why is there a local pressure bump  near $r_H$ = 150 AU (c.f.~Figure \ref{Fig6}), and not a continuous rise towards smaller heliocentric distances?  Our models give several clues.  First, consider the unlikely case  of a totally sealed comet surface (the ``steam cooker'' approximation). Figure (\ref{Fig10}) shows that, as expected,  the calculated pressure  closely follows the temperature  of the ice with no pressure bump  and independent of the adopted physical parameters of the mantle. This numerical experiment gives the total cumulative CO gas production on the whole path, mainly produced after the CO Mendis point. The steam cooker approximation proves that the  bump and changes inward of the Mendis point are products of gas transport processes in the mantle because these fall to zero when the surface is sealed.  Second, apart from this pathological case, which we do not expect to find in nature, the pressure bump exists in all models, albeit with widely different amplitudes depending on the model parameters, even in the case of an almost totally (99\%) permeable mantle. \\

\cite{2017MNRAS.469S.217P} noted that the uppermost  surface layer of the mantle may be less permeable to flow than the deeper layers. We model  this by reducing the comet surface pore radii, $r_{s}$, compared to the mantle pore radii, $r_{p}$. Figure (\ref{Fig7}) shows that the flux $J_{CO}$ amplitude and behavior is strongly controlled by the assumed structural properties of  the comet mantle. The more the flux is inhibited, the higher is the flux bump. The location of the Mendis point at the junction between the two thermal regimes was noted above. It is at this confluence point that the lower flow entering the mantle, ($x=B$ in Figure \ref{Fig1}) stops increasing over time and slowly becomes constant in response to the flattening of the temperature dependence caused by strong CO sublimation in Regime 2. This happens while, at the surface of the comet ($x=A$ in Figure \ref{Fig1}), and also in the mantle depending on the pore radius, the gas pressure remains high. The transition from the rising ice front temperature in Regime 1 to the nearly constant ice front temperature in Regime 2 leads to a temporary under-supply of CO at the base of the mantle, relative to the loss rate from the surface, creating a local maximum as in Figure (\ref{Fig6}) or less visible but there in the orange curve of Figure (\ref{Fig7}). The width of the bump is a measure of the mantle's holding or response time. The under-supply of CO starting around 150 AU,  relative to the rate of loss from the surface, is responsible for the peak as the mantle switches from regime 1 to regime 2. 
\\

\subsection{Retention Time Scale }
\label{retention-time}

The porous mantle of the comet acts as a buffer to the gas flow produced by sublimation at the bottom. In order to understand this buffer and explore the physical parameters that affect the gas retention time, we estimate in this section the time scale corresponding to this buffer. \\

We first define the Knudsen Number $K_n$ by

\begin{equation}
K_n~ =~ \frac{\lambda}{r_{p}}\label{key6bis2}
\end{equation}

\noindent where $r_p$ is the pore radius and the mean free path is given by

\begin{equation}
\lambda = \frac{1}{\sqrt{2}\pi {\sigma}^{2} n}\label{key6bis3}.
\end{equation}
 
 \noindent Here, $\sigma$ is the effective particle radius and $n$ is the number density in the gas.  The Knudsen number gives a measure of the  importance of gas - gas collisions relative to gas - particle collisions with the solid material making up the porous mantle.

We  define an effective flow velocity of the CO gas through the mantle, $V$, from $V = J_{CO}/n$, where  $n$ is the gas number density.  Then, if $\Delta L$ is the mantle thickness, Equation (\ref{key6bis1}) can be rewritten as  


\begin{equation}
V ~~=~~\frac{\epsilon D_{K}}{n}\frac{\Delta n}{\Delta L}+\frac{\epsilon \kappa}{\mu} \frac{\Delta p}{\Delta L}\label{key6bis5}
\end{equation}

The time taken for the flow to cross $\Delta L$ defines a holding or retention timescale, $\tau= \frac{\Delta L}{V}$.   Equation (\ref{key6bis5}) gives 

\begin{equation}
\tau = \frac{{\mu n \Delta L}^{2}} {\epsilon \kappa n \Delta p(n,T) + \epsilon D_{K} \mu \Delta n }\label{key6bis6} 
\end{equation}

\noindent Here $\Delta p = (p_{A}-p_{B})$ and $\Delta n = (n_{A}-n_{B})$ are the differences in pressure and gas density, respectively, between the surface of the comet (level A in Figure \ref{Fig1}) and the bottom of the  mantle (level B).   For simplicity, we calculate  the intermediate coefficients $D_{K}$, $\mu$ and $\lambda$ of this equation,  using the averaged value of temperature, molecular density and pressure (i.e.~$T_{m}= \frac{T_{A}+T_{B}}{2}$). \\

\begin{figure}[ht!]
	\plotone{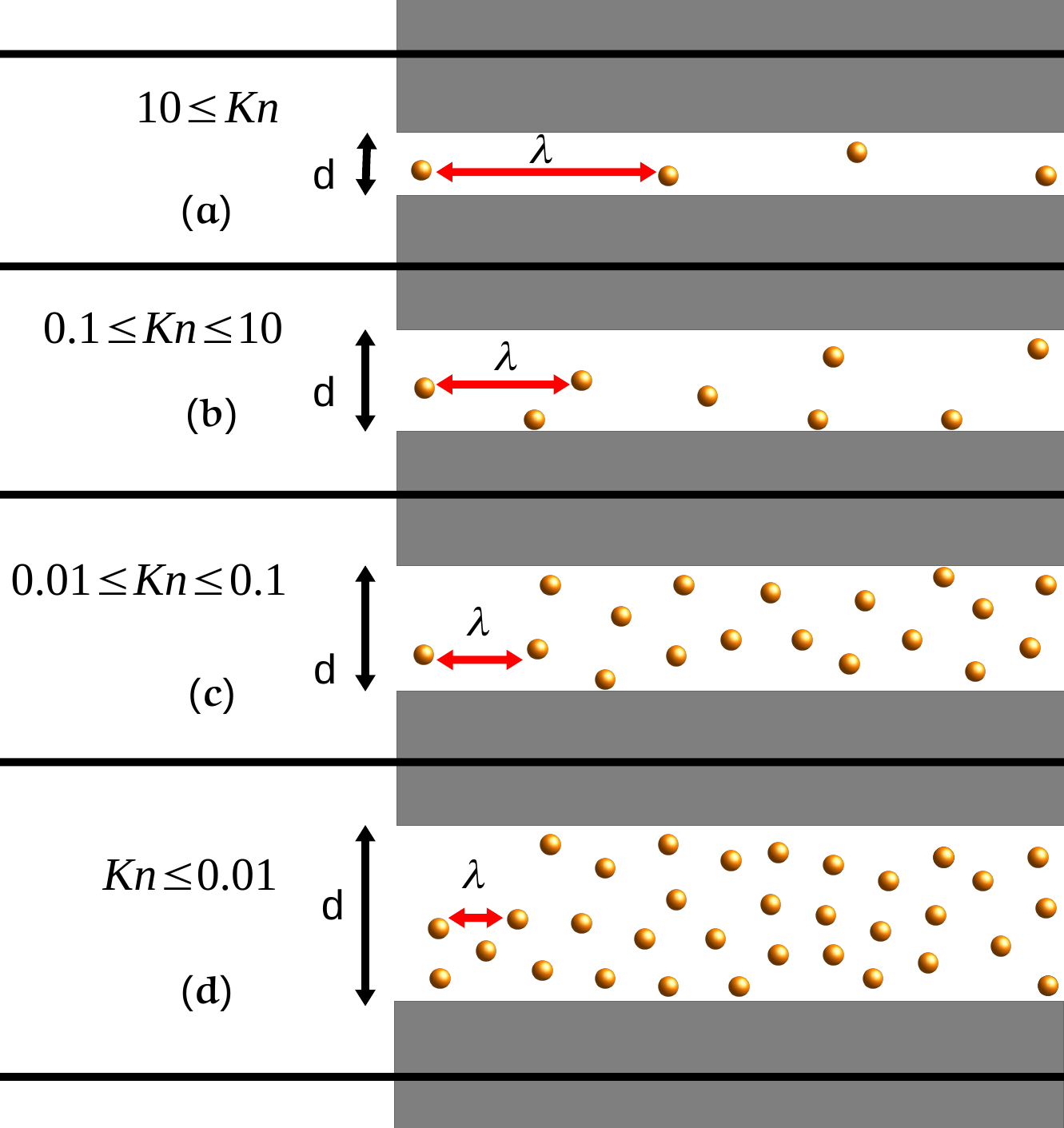}
	\caption{ 
	The diagram shows how the different gas transport processes should evolve as  the comet moves from the Oort cloud to the inner solar system,  starting from high $Kn$ numbers ($Kn=\lambda/d$, where $\lambda$ is the mean free path and $d$ is the pore size) to low $Kn$ numbers and changing continuously over a wide range. Flow regimes evolve from  Knudsen free diffusion (a), followed by a transition regime (b)  that turns into a slip regime (c)  that leads to  a viscous continuum regime (d). Figure modified from \citep{moghaddam2016slip} \& from \url{https://tinyurl.com/y3xyj9ss}. \label{Fig11}}
\end{figure}

Equation (\ref{key6bis6}) shows that $\tau$ is governed by the two main modes of gas transport, namely 1) Knudsen diffusion in the second denominator term and 2) viscous flow (first denominator term). Figure (\ref{Fig11}) illustrates these two modes as well as the intermediate transition and slip stages that are intermediate steps between them over the wide range of pressures and pore sizes. The Knudsen number approximately indicates the ranges of the different modes.  \\

\begin{figure}[ht!]
	\plotone{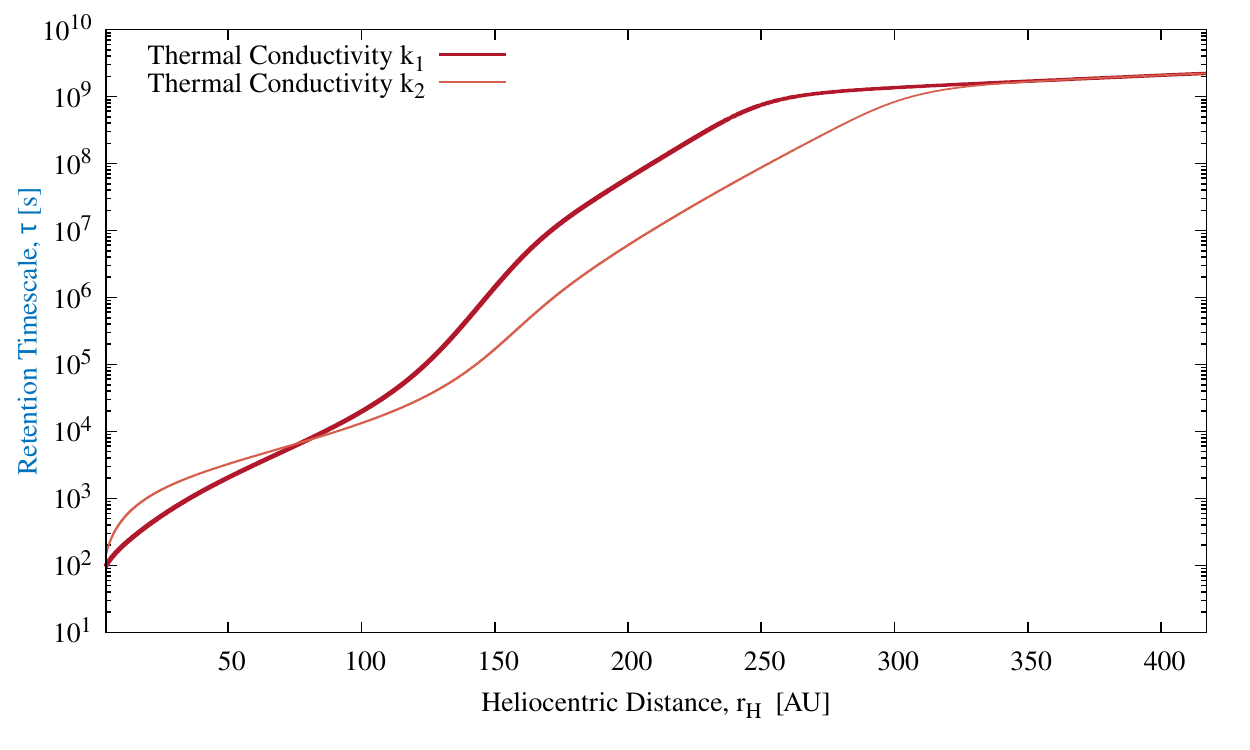}
	\caption{ 
	 Retention time scale $\tau$ (Eq. \ref{key6bis6}) vs. the heliocentric distance $r_{H}$ for two thermal conductivities (in W m$^{-1}$ K$^{-1}$); (1) $k_1(T)= 10^{-2}+10^{-2} (4 {\sigma}_{IR} T^{3})$, (2)  $k_2(T)= 10^{-4}+10^{-2} (4 {\sigma}_{IR} T^{3})$  
 \label{Fig12}}
\end{figure}

\begin{figure}[ht!]
	\plotone{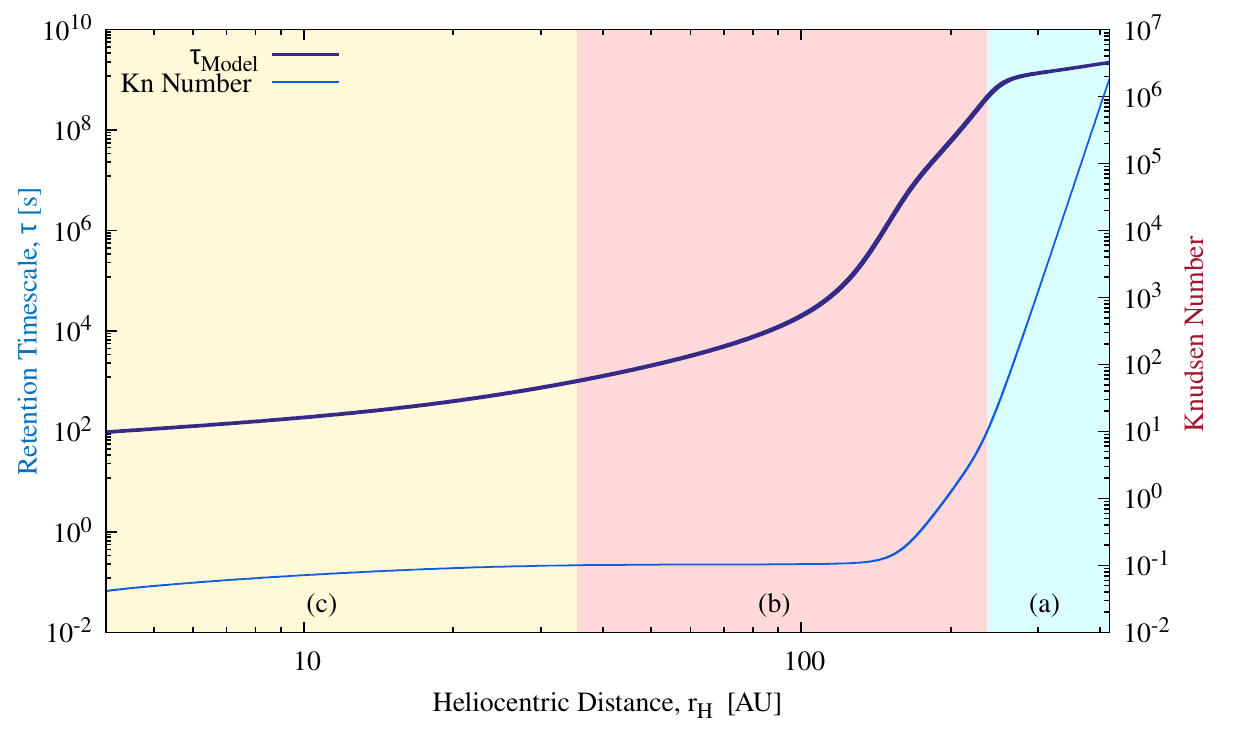}
	\caption{ 
	This figure shows retention time  ($\tau_{Model}$) (thick line) in various transport modes. The Knudsen number (thin line) is shown on the right- hand axis. (a) is the Knudsen Diffusion region ($10 \leq Kn$), (b)  the Transition regime  ($0.1 \leq Kn \leq 10$) and,  (c)  the Slip regime region where ($ 0.01 \leq Kn \leq 0.1$).  We have assumed for purposes of illustration a thermal conductivity $k(T)= 10^{-2}+10^{-2} (4 {\sigma}_{IR} T^{3})$ (W m$^{-1}$ K$^{-1}$), mantle pore radius $r_{p}= 10^{-4}$ m and, comet surface pores are taken equal to 95\% $r_{p}$.	
	\label{Fig13}}
\end{figure}

At what heliocentric distance does the Knudsen-Viscous transition occur? The answer will depend on the structure of the dust mantle, in particular the pore radius and the average free path, $\lambda$, which varies with pressure. Figure (\ref {Fig12}) shows how the retention time changes with the heliocentric distance, $r_H$, focusing on the coupling between the Knudsen flow and the viscous flow and the different steps between the two flows.  It reveals two very distinct configurations. At large $r_H$ we find an almost constant time scale of one century, leading to a slow and almost uniform motion of the gas leaving the comet surface at a global bulk flow velocity of V = $\frac{1}{3}~ 10^{-9} $ (m $s^{-1}$). The retention time scale  decreases dramatically as the comet approaches the sun, falling to about an hour at 50 AU.  This indicates an acceleration of the gas, which we approximate by estimating the slope of the Figure (\ref{Fig12}), given the assumed physical parameters.  By examining the behavior of the retention time Figure (\ref{Fig13}), we can easily identify the four known modes for each independently computed Knudsen number value and this is then a check of the model. The model was able to capture the complete physics of the flow without any artificial intervention by using the Knudsen number to switch from one mode to the other.\\ 

\begin{figure}[ht!]
	\plotone{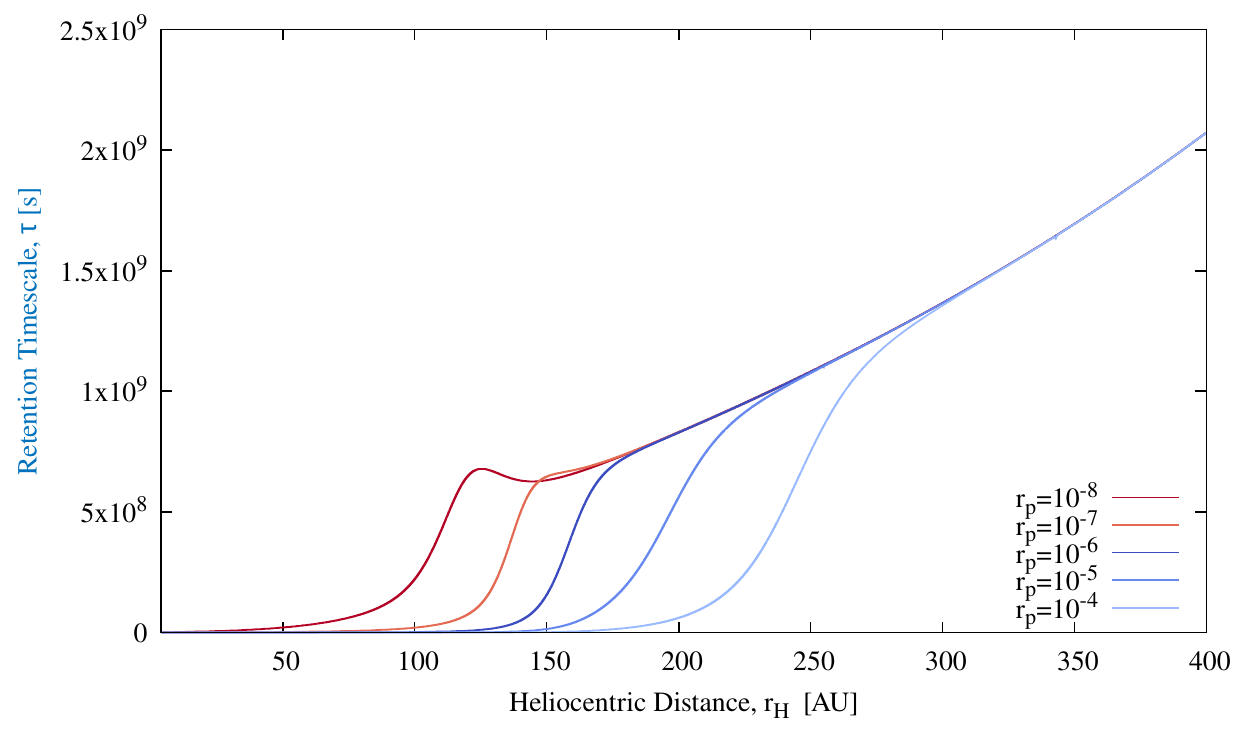}
	\caption{ The effect of pore radius on retention time (Equation \ref{key6bis6}).  The curves show that the behavior of $\tau$ changes depending on whether the transition (Knudsen-Viscous) in the gas transport occurs before (the blue curves) or after (the red curves) the 150 AU Mendis point. It is shown here that the sublimation temperature transition has an important retention effect and this effect scales with $r^{2}$.  We have assumed  that $k(T)= 10^{-2}+ 10^{-2}(4 {\sigma}_{IR} T^{3})$ (W m$^{-1}$ K$^{-1}$), with mantle thickness 1 m and  surface pore radii $r_s = 0.95 r_m$. \label{Fig14}}
\end{figure}

Knudsen diffusion predominates mainly at low pressure and the model allows a viscous flow to develop continuously when the pressure becomes appreciable. As we produce gas by sublimation, Figures (\ref{Fig14}) show that depending on the transport mode, the duration of this retention time can vary widely.    \\

\subsection{Largest ejected dust particle }


To be ejected, a dust particle must first overcome cohesive forces, $S_0$, exerted on it by the surrounding dust material \citep{2015A&A...583A..12G}. The minimum size of grains that could escape is \citep{2019AJ....157...65J}:

\begin{equation}\label{key12}
      a_{s}(r_{H}) =\frac{S_{0}}{C_{D}f_s(r_{H})V_s(r_{H})},
\end{equation}

\noindent where $C_D$ is a dimensionless constant of order unity, $f_s$ (kg m$^{-2}$ s$^{-1}$) is the mass sublimation flux and $V_s$ is the speed of the sublimated gas.  Once detached, the dust grain needs a force to accelerate above escape velocity. This force can be provided by gas drag. For  particles that are spherical and of uniform density, the maximum size of grains that can be ejected is approximated by \citep{2019AJ....157...65J}:

\begin{equation}\label{key14}
      a_{c}(r_{H}) =\frac{9C_{D}f_s(r_{H})V_s(r_{H})}{16 \pi G\rho\rho_{n}r_{n}}.
\end{equation}
 
\noindent Here $a_{s}(r_{H})$ \& $a_{c}(r_{H})$ are the dust grain sizes escaping the comet surface  against cohesive forces  and against  cometary gravity, respectively. The model outputs : $f_s(CO)$ and $V_s(CO)=\sqrt{\frac{8k_{B}T_s}{\pi \mu m_H}}$ are respectively the CO gas outflow and its thermal speed, taken at the comet surface, where $T_s$ is the comet surface temperature, $\mu$ is the molecular weight, $m_H$ the mass of hydrogen,  $G$ is the gravitational constant, $\rho=\rho_{n}$ respectively dust and nucleus densities, $C_{D}$ is the dimensionless drag efficiency coefficient, assumed to be equal to 1,  cohesive strength constant $S_{0}\approx 1.5 \times 10^{-4}$ N m$^{-1}$ \citep{2014M&PS...49..788S}, and  $r_{n}$ is the comet nucleus radius. \\

\begin{figure}[ht!]
	\plotone{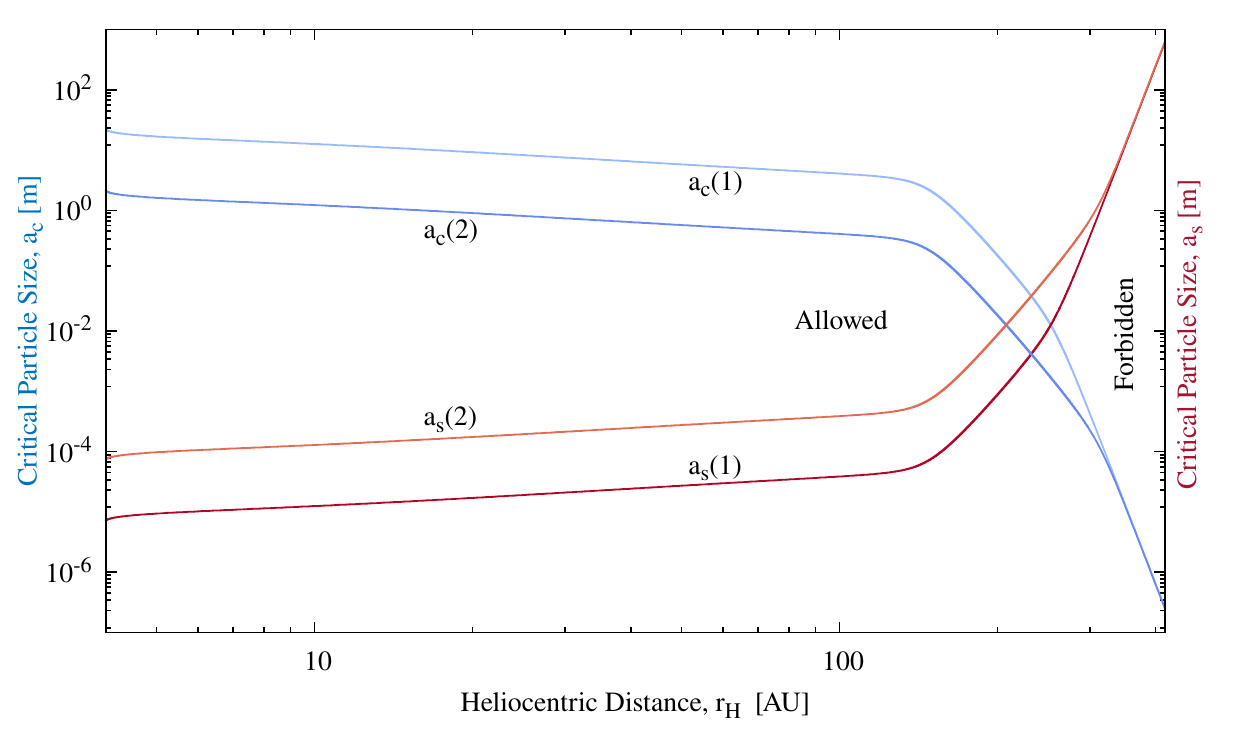}
	\caption{ The largest dust particle escaping the comet surface. The forbidden and the allowed zones for dust particle production are separated by the bottleneck intersection point. The critical sizes $a_{c}$ \& $a_{s}$ are labeled (1) \& (2) to show the effect of mantle pore sizes (1) $r_p$ is $10^{-4}$ m and (2) $r_p$ is $10^{-3}$ m, here thermal conductivity is $k(T)= 10^{-4}$ (W m$^{-1}$ K$^{-1}$). We observe on this figure that for $r_p= 10^{-3}$ (m), we can't overcome cohesion and detach 0.1 mm dust particles (see the inner curves $a_{c}(2)$ \& $a_{s}(2)$) unless we reach a heliocentric distance of $\sim$150 AU. The assumed thickness of the mantle  is 1 m and the surface pores have radii $r_s = 0.97 r_p$. \label{Fig15}}
\end{figure}

Figure (\ref{Fig15}) displays Equations (\ref{key12}) and (\ref{key14}) as a function of $r_H$. Particle radii must lie above the red curve in order for gas drag to exceed inter-grain cohesive forces and below the blue curve for gas drag to exceed the gravitational pull of the nucleus.   The Figure shows that these conditions can be simultaneously satisfied at all distances inside the $r_H \sim$ 150 AU CO Mendis point, thus presenting a resolution of the cohesion bottleneck problem for all active comets observed to-date.

\clearpage

\section{Summary}

We present a detailed gas-physics model of sublimation of cometary CO ice  buried beneath a permeable, refractory dust mantle.   The model is solved as a function of time and distance along the orbit of the distantly active long-period comet C/2017 K2.

\begin{enumerate}

\item Our main result is that we find an unexpected local peak in CO production  at very large heliocentric distances ($r_H \approx~150$ AU), caused by the build up of pressure beneath a permeable mantle of modest thickness.   

\item This peak, whose magnitude is a function of several mantle microphysical parameters, corresponds to the Mendis point, where the buried CO ice front  transitions between distinct temperature regimes.

\item While modeled on C/2017 K2 as a specific case, our conclusions are general.  Comets entering the solar system from Oort cloud and interstellar distances may become active far beyond the planets, at heliocentric distances $r_h \sim$ 150 AU.  Observational attempts to detect such ultra-distant activity are encouraged.

\end{enumerate}

\begin{acknowledgments}
We thank the anonymous referee for their valuable comments.   We thank B. Schmitt and N. Fray for providing the empirical pressure-temperature data for cometary ices. 
\end{acknowledgments}
\newpage

\appendix

To justify our neglect of conduction into the cold core beneath the sublimation front (Equation \ref{key3}), we can show that sublimation dominates the energy budget over conduction down into the cold core. If $l$ is the thermal skin depth to the cold core, then,

\begin{equation}\label{A1}
l \sim \left(\frac{k \tau}{\rho c_p} \right)^{1/2}.
\end{equation}

Here, $k$ is the thermal conductivity of the material,  $\rho$ and $c_{p}$ are respectively the  density and  heat capacity.
Quantity $\tau$ is the heating timescale. \\

The conducted heat flux  into the cold core is $F_{C}= k (d T/d x) \approx k\Delta T /l $, where $\Delta T = T_{CO} - T_C$ is the temperature difference between the top and bottom of the heated layer of thickness $l$.  Then

\begin{equation}\label{A3}
F_{C}= \left(\frac{k \rho c_p }{\tau}\right)^{1/2} \Delta T
\end{equation}

For comparison, the energy flux used by sublimation is

\begin{equation}\label{A4} 
F_{sub}=f_{CO} L_{CO}
\end{equation}

\noindent where $f_{CO}$ is the mass flux of CO in kg m$^{-2}$ s$^{-1}$ and $L =2.66\times 10^{5}$ J kg$^{-1}$ is the latent heat of CO from Table (\ref{tab1}). \\

Conduction into the cold core is smaller than sublimation when $F_{C} < F_{sub}$, leading to 

\begin{equation}\label{A6}
f_{CO} >  \left(\frac{k \rho c_p }{\tau}\right)^{1/2} \frac{\Delta T}{L_{CO}}.
\end{equation}

\noindent Setting $\Delta T = (26 - 10) = 16$ K, $k = 10^{-3}$ W m$^{-1}$ K$^{-1}$, $\rho$ = 533 kg m$^{-3}$, $c_p$ = 119 J kg$^{-1}$ K$^{-1}$ and $\tau \approx 5 \times 10^{9}$ s as the travel time from the Mendis point to 10 AU, we find that sublimation dominates provided
$f_{CO} > 6.7 \times 10^{-9}$ kg m$^{-2}$ s$^{-1}$.  For comparison, Eq. (\ref{key9}) gives  $f_{CO}=1.4 \times 10^{-7}$ kg m$^{-2}$ s$^{-1}$ at $T$ = 26 K confirming that the  sublimation energy flux is much larger than the conducted energy flux.

\newpage



\clearpage

\begin{deluxetable*}{lllc}

 	\tablenum{1}
 	\tablecaption{Physical Parameters  \label{tab1}}
 	\tablewidth{0pt}
 	\tablehead{
 		\colhead{Parameters}& \colhead{Symbols} &
 		\colhead{Values} & \colhead{References} \\
 	}
 	\startdata
 	Oort Cloud Temperature  (K)        & $T_{OC}$  & $10$  & \cite{2018AJ....156..243B} \\
    Gravitational Constant ($ m^{3} kg^{-1} s^{-2}$ )         & $G$ & $6.67\times 10^{-11} $  &   \\
 	Comet Nucleus Surface Albedo        & $A_{B}$  & $0.04$  & \\
 	Comet Nucleus Radius  (m) & $r_{n}$ & $9\times 10^{3}$ & \cite{2017ApJ...847L..19J}\\
 	Comet Nucleus Density (kg  m$^{-3}$) & $\rho_{n}$ & $533$ ~ &  \\
 	Dust Grains Density (kg  m$^{-3}$) & $ \rho=\rho_{n}$ & $533$  &\\
 	Porous Crust Thickness (m) & $L$ & $10^{-2}$ -- $10^{2}$  & \\
 	Main Pore Radius (m) & $r_{p}$ & $10^{-8}$ -- $10^{-2}$ & \\
 	Porosity  & $\epsilon$ & 0.80 & 	 \\
 	Emissivity & $\epsilon_{IR}$ & $0.95$ &  \\
 	Stefan-Boltzmann Constant (W m$^{-2}$ K$^{-4}$ )& $\sigma$ & $5.67\times10^{-8}$ &  \\
 	Solar Constant (W m$^{-2}$ )         & $F_{\odot}$ & $1360.8$  &   \\
 	Boltzmann Constant (J  K$^{-1}$ )         & $k_{B}$ & $1.380649 \times10^{-23} $  &   \\
 	Ideal Gas Constant (J  K$^{-1}$  mol$^{-1}$ )         & $R$ & $8.314 $  &   \\
 	Latent Heat--CO (J mol$^{-1}$)& $E_{CO}$ & $7.45 \times 10^{3}$  & Data \& Fits Fig. (\ref{Fig2}) \\ 
 	Sublimation Pressure of CO (N m$^{-2}$)& $P(T)$ & $8 \times 10^{9}e^{-7450/RT}$   & Data \& Fits Fig. (\ref{Fig2}) \\ 
 	Thermal Conductivity (W m$^{-1}$ K$^{-1}$)& $k(T)= A+B(4\sigma \epsilon_{IR}T^{3})$  & $A$ \& $B =10^{-4}$-$10^{-1}$  &  \cite{1977Moon...17..359M} \\ 
 	  \enddata
 \end{deluxetable*}

\clearpage

\clearpage

\clearpage

\clearpage

\clearpage

\clearpage

\clearpage

\clearpage

\clearpage

\clearpage

\clearpage

\clearpage

\clearpage

\clearpage

\clearpage

\clearpage

\clearpage	

\clearpage

\end{document}